%% file: paper.tex
\documentclass[]{bytedance_seed}

\usepackage[toc,page,header]{appendix}

\usepackage{minitoc}
\usepackage{colortbl}
\usepackage{booktabs}
\usepackage{hhline}
\usepackage{amsmath}
\usepackage{amssymb}
\usepackage{wrapfig}
\usepackage{fix-cm}

\makeatletter
\def\shline{\noalign{\ifnum0=`}\fi\hrule \@height 1pt \futurelet\reserved@a\@xhline}
\makeatother

\title{ WavTTS: Towards High-Quality Zero-Shot TTS via Direct Raw Waveform Modeling }

\author[1,2]{Wenxi Chen}
\author[3]{Dongya Jia}
\author[1,2]{Yushen Chen}
\author[1,2]{Zhikang Niu}
\author[1,2]{Yuzhe Liang}
\author[1]{Xiquan Li}
\author[1]{Ruiqi Yan}
\author[1,2]{Ziyang Ma}
\author[1,2]{Guanrou Yang}
\author[3]{Sanyuan Chen}
\author[3]{Yue Wang}
\author[3]{Zhuo Chen}
\author[1]{Kai Yu}
\author[1,2\dagger]{Xie Chen}

\affiliation[1]{Shanghai Jiao Tong University}
\affiliation[2]{Shanghai Innovation Institute}
\affiliation[3]{ByteDance Seed}

\contribution[\dagger]{Corresponding author}

\abstract{
Recently, diffusion models operating on VAE latents or mel-spectrograms have become the dominant paradigm for zero-shot TTS. Although these compressed representations improve generation efficiency, they inevitably suffer from information loss and non-end-to-end training. Theoretically, directly modeling raw waveforms circumvents these issues; however, this direction remains underexplored and is often deemed difficult due to the extremely long sequence length of audio signals. To overcome this, we propose \textbf{WavTTS}, the first raw waveform generative TTS model that substantially narrows the gap with latent-space generative models. Built upon the flow matching with Diffusion Transformer (DiT), WavTTS directly models speech waveforms via a simple patchification strategy, while integrating multi-scale mel-spectrogram supervision to provide perceptual guidance during training. Furthermore, we investigate the impact of prediction targets and noise scheduling in waveform diffusion, and develop an effective schedule design to improve generation quality. Evaluations on open-source benchmarks demonstrate that WavTTS closely approaches the performance of current state-of-the-art latent generative zero-shot TTS models, while substantially outperforming previous end-to-end speech generation models. Our findings demonstrate the feasibility of scaling diffusion-based TTS directly in the waveform space, opening a new direction for end-to-end speech generation. 
}

\correspondence{Xie Chen at \email{chenxie95@sjtu.edu.cn}}

\checkdata[Project Page]{\url{https://wavtts.github.io}}
\checkdata[Code and Model]{\url{https://github.com/cwx-worst-one/WavTTS}}

\begin{document}
\maketitle


\input{sections/introduction}
\input{sections/relatedwork}

\input{sections/approach}
\input{sections/experiments_setup}
\input{sections/main_results}
\input{sections/ablation_study}
\input{sections/conclusion}

\clearpage

\bibliographystyle{plainnat}
\bibliography{main}

\clearpage

\beginappendix

\input{sections/appendix}

\end{document}

%% file: sections/introduction.tex
\section{Introduction}

Recent years have witnessed remarkable progress in text-to-speech (TTS)~\citep{FastSpeech, FastSpeech2, NaturalSpeech, NaturalSpeech2}, where current zero-shot TTS models are capable of achieving voice cloning and high-quality speech generation given only a brief audio prompt~\citep{VALLE, zhang2023speak, Seed-TTS}. 
Existing architectures predominantly fall into autoregressive (AR)~\citep{FireRedTTS, FireRedTTS-2, MOSS-TTS, DiTAR, Qwen3-TTS, VoxCPM, IndexTTS2} and non-autoregressive (NAR)~\citep{Voicebox, ZipVoice, M3-TTS, LongCat-AudioDiT, OmniVoice, F5R-TTS} paradigms. 
AR models based on next-token prediction are highly expressive and eliminate the need for explicit duration predictors, but they are constrained by heavily compressed and quantized discrete tokens~\citep{VALLE2, VALLE-R, ELLA-V, VOICECRAFT}, while also suffering from high inference latency and exposure bias. 
In contrast, NAR models, primarily built upon diffusion-based architectures~\citep{ddpm, score}, greatly improve generation speed through parallel inference in continuous acoustic spaces. Recent advances further remove the need for external duration predictors through implicit text-to-representation alignment~\citep{E2-TTS, F5-TTS, DiTTo-TTS}. 
However, whether utilizing highly compressed VAE latents or mel-spectrograms that discard phase and high-frequency details, these continuous representations remain inherently lossy, imposing an upper bound on generation quality. 
Furthermore, the conventional two-stage training paradigm, which relies on pre-trained autoencoders or vocoders, inevitably introduces accumulated errors and decoding artifacts. 
This motivates us to revisit the existing speech generation pipeline: \textit{Can we achieve high-quality zero-shot TTS by directly modeling the raw, uncompressed waveform space?}

Indeed, several early studies~\citep{VITS, EATS} have explored end-to-end speech generation without relying on lossy acoustic intermediates. 
WaveNet~\citep{WaveNet} pioneered neural raw waveform generation by autoregressively predicting audio samples, but its practical use was severely limited by prohibitively slow inference. 
Although subsequent studies improved waveform generation efficiency through parallel generation~\citep{ParallelWaveNet, Clarinet}, block-wise modeling~\citep{Wave-tacotron}, or diffusion-based refinement~\citep{E3TTS, DiffAR}, raw waveform TTS remains highly challenging. 
The extremely high temporal resolution of raw audio requires models to capture long-range linguistic dependencies while preserving fine-grained phase, periodicity, and high-frequency structures within a high-dimensional continuous space. 
Moreover, prior waveform-based TTS systems have rarely been scaled to modern zero-shot settings with in-context speaker prompting, leaving a substantial generalization gap compared with recent mel- or latent-space generative TTS systems.

In this paper, we revisit waveform-space generative modeling and propose \textbf{WavTTS}, a high-quality, end-to-end zero-shot TTS model. 
Based on flow matching~\citep{FlowMatching} with Diffusion Transformers (DiT)~\citep{DiT}, WavTTS enables truly end-to-end speech generation by eliminating the reliance on pre-trained autoencoders or vocoders, as illustrated in Figure~\ref{fig:modeling}. 
To address the computational challenges posed by extremely long raw waveform sequences, we employ a simple non-overlapping patchification strategy.
Furthermore, to improve optimization efficiency, we adopt an $x$-prediction formulation~\citep{JiT}, which directly predicts the clean waveform from noisy inputs. This formulation naturally allows us to incorporate multi-scale mel-spectrogram supervision, providing perceptual guidance that accelerates convergence and improves generation quality.

Building upon this architecture, we further reveal the critical role of noise design in waveform-space flow matching. 
By aligning signal-noise variances and shifting the temporal schedules toward high-noise regimes during both training and inference, we substantially enhance model robustness, speech naturalness, and intelligibility.
Finally, our scaling analysis demonstrates that large-scale data and matched model capacities are essential for unlocking the potential of high-dimensional waveform modeling. 
Comparisons with alternative lossless representations, such as STFT and MDCT, further demonstrate the simplicity and effectiveness of direct time-domain generation.
In summary, our main contributions are as follows:

\begin{itemize}
    \item We propose \textbf{WavTTS}, a flow-matching framework that performs end-to-end zero-shot TTS directly in the waveform space. This framework eliminates the reliance on pre-trained autoencoders, neural codecs, or vocoders, thereby simplifying the speech generation pipeline.
    
    \item We introduce key designs tailored for effective waveform-space generation, including waveform patchification, an $x$-prediction objective coupled with multi-scale mel-spectrogram supervision, and signal-noise variance alignment paired with noise-shifted temporal schedules across both training and inference.
    
    \item To the best of our knowledge, WavTTS is the first raw waveform generative TTS system to closely approach the performance of mainstream state-of-the-art NAR zero-shot TTS models. This validates the feasibility of direct waveform modeling and challenges the prevailing assumption that high-quality TTS necessarily requires intermediate acoustic features or discrete tokens.
\end{itemize}

%% file: sections/relatedwork.tex
\section{Related Work}

\begin{figure*}[t]
    \centering
    \includegraphics[width=\textwidth]{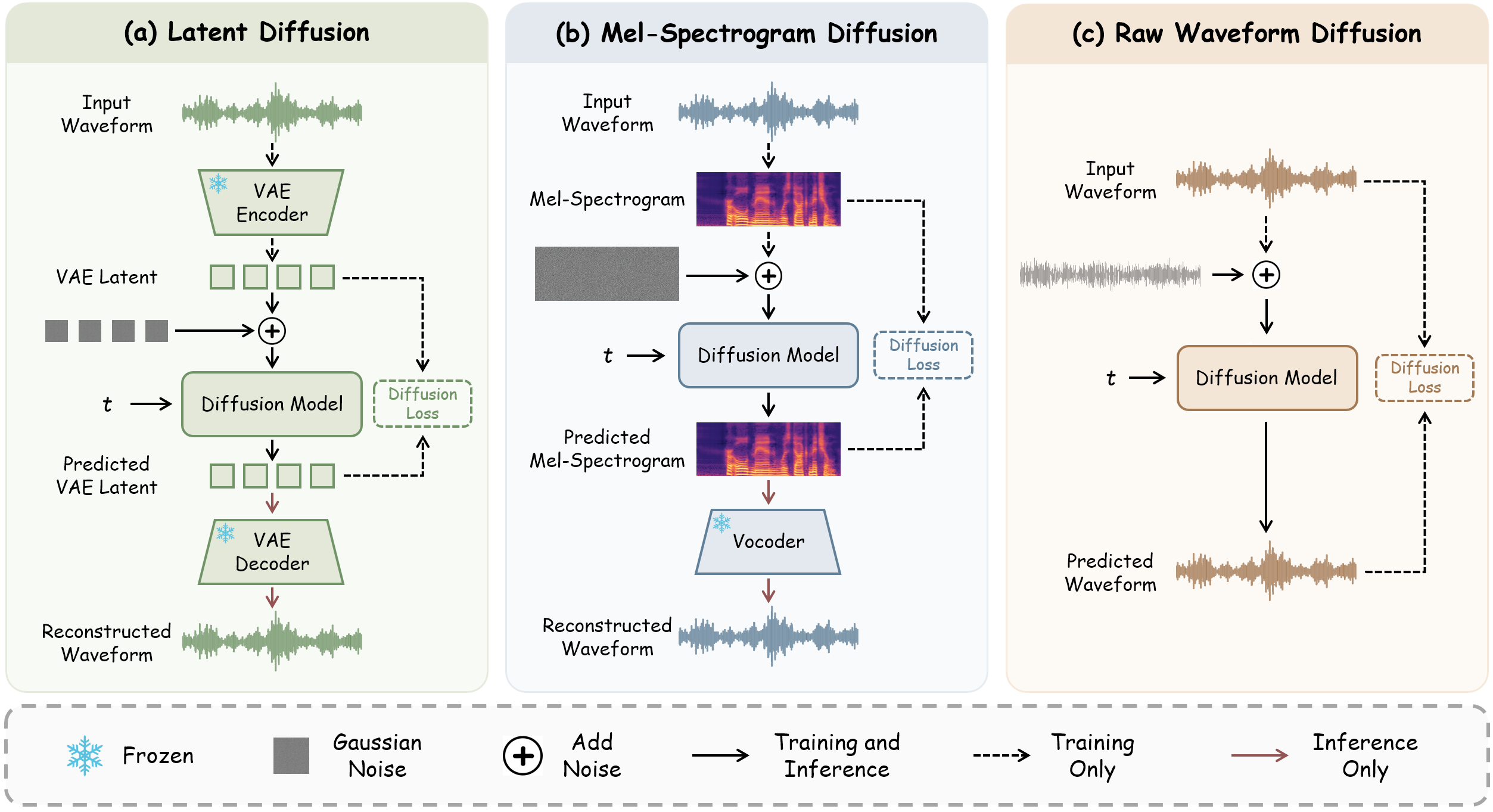}
    \vspace{-5mm}
    \caption{Diffusion paradigms across different representation spaces in text-to-speech synthesis. (a) \textbf{Latent Diffusion} modeling on highly compressed VAE representations. (b) \textbf{Mel-Spectrogram Diffusion} modeling on spectrograms with discarded phase information. (c) \textbf{Raw Waveform Diffusion} modeling directly on lossless audio waveforms.}
    \label{fig:modeling}
    \vspace{-3mm}
\end{figure*}

\textbf{Diffusion-based TTS.} 
Diffusion-based generative models have emerged as a dominant paradigm for NAR speech synthesis~\citep{NaturalSpeech2, NaturalSpeech3}. Early approaches, such as Diff-TTS~\citep{diff-tts}, Grad-TTS~\citep{grad-tts}, and ProDiff~\citep{prodiff}, primarily relied on denoising diffusion probabilistic models (DDPMs) with score-matching objectives~\citep{ddpm, score}. 
More recent studies have shifted toward flow matching~\citep{FlowMatching, Matcha-TTS, DiTTo-TTS, VoiceFlow}, where representative systems such as Voicebox~\citep{Voicebox} and Matcha-TTS~\citep{Matcha-TTS} leverage optimal transport and continuous-time flows to improve generation quality while enabling efficient ODE-based sampling.
In terms of acoustic representation, most existing diffusion-based TTS models operate in compressed continuous spaces, specifically VAE latents~\citep{M3-TTS, LongCat-AudioDiT, MegaTTS3, VibeVoice} (Figure~\ref{fig:modeling}(a)) and mel-spectrograms~\citep{F5-TTS, E2-TTS, ZipVoice} (Figure~\ref{fig:modeling}(b)). 
These representations substantially reduce sequence length and computational cost.  
However, they are inherently lossy and rely on pre-trained autoencoders~\citep{VAE, Semantic-VAE} or vocoders~\citep{Vocos, BigVGAN}, leading to a multi-stage generation pipeline that may introduce compounding errors and reconstruction artifacts.
This motivates us to revisit end-to-end raw waveform generation under modern diffusion-based TTS frameworks, as illustrated in Figure~\ref{fig:modeling}(c), aiming to eliminate lossy intermediate acoustic representations while retaining the efficiency and scalability of NAR generation.

\textbf{Raw Waveform Modeling.}
Despite the computational challenges of high temporal resolution, direct raw waveform modeling remains highly appealing. WaveNet~\citep{WaveNet} pioneered autoregressive modeling on raw waveforms.
Subsequent works~\citep{ParallelWaveNet, Diffwave, FastDiff} substantially improved generation efficiency but mainly served as vocoders conditioned on intermediate acoustic features (e.g., mel-spectrograms) rather than full TTS systems. 
Early end-to-end TTS efforts explored various architectures with knowledge distillation, GANs, normalizing flows, or diffusion models~\citep{Clarinet, FastSpeech2, EATS, Wave-tacotron, WaveGrad2}. 
Notably, while VITS~\citep{VITS} and JETS~\citep{JETS} achieve high-quality end-to-end training, they still rely on intermediate features, latent variable modeling, or adversarial vocoders.
More recently, diffusion-based native waveform generation has gained increasing attention. 
DiffAR~\citep{DiffAR} applies diffusion in waveform space but is limited by slow autoregressive generation. 
E3-TTS~\citep{E3TTS} proposes a simple NAR diffusion approach for waveform generation; however, it lacks systematic comparisons with mainstream NAR baselines under large-scale controlled settings.
Concurrent work WavFlow~\citep{WavFlow} also investigates direct waveform generation, but it focuses on sound effects generation rather than text-to-speech synthesis.
This gap highlights the need for a scalable waveform generative TTS system that eliminates intermediate representations while enabling efficient and high-quality generation in the high-dimensional time-domain space.

\textbf{Insights from Pixel-Space Diffusion.}
In computer vision, the exploration of diffusion models in high-dimensional pixel space~\citep{ADM, song2019generative, ddpm, Improveddiffusion} predates their latent-space counterparts. 
Early architectures~\citep{simplediffusion} typically relied on U-Nets~\citep{U-Net} with dense convolutions and long residual connections, which are computationally prohibitive and severely limit model scaling. 
To alleviate this burden, subsequent studies explored various directions, such as decomposing the diffusion process across multiple resolution scales~\citep{PixelFlow, RelayDiffusion}, integrating autoregressive architectures with normalizing flows~\citep{TarFlow, FARMER}, or introducing auxiliary decoders to recover high-frequency details~\citep{DeCo, DiP, PixelDiT, PixNerd}.
Alongside the rise of patch-based architectures, recent works have also re-examined diffusion objectives for high-dimensional generation. 
JiT~\citep{JiT} proposes directly predicting the clean image, i.e., $x$-prediction, to improve modeling in pixel space, while PixelGen~\citep{PixelGen} further incorporates perceptual loss to better capture the perceptual manifold of pixels. 
These advances in visual generation inspire our waveform modeling design: we adopt an $x$-prediction objective and combine it with multi-scale mel-spectrogram supervision as a perceptual training signal, enabling high-quality raw waveform generation.

%% file: sections/approach.tex
\section{WavTTS}
\begin{figure}[t]
    \centering
    \includegraphics[width=\linewidth]{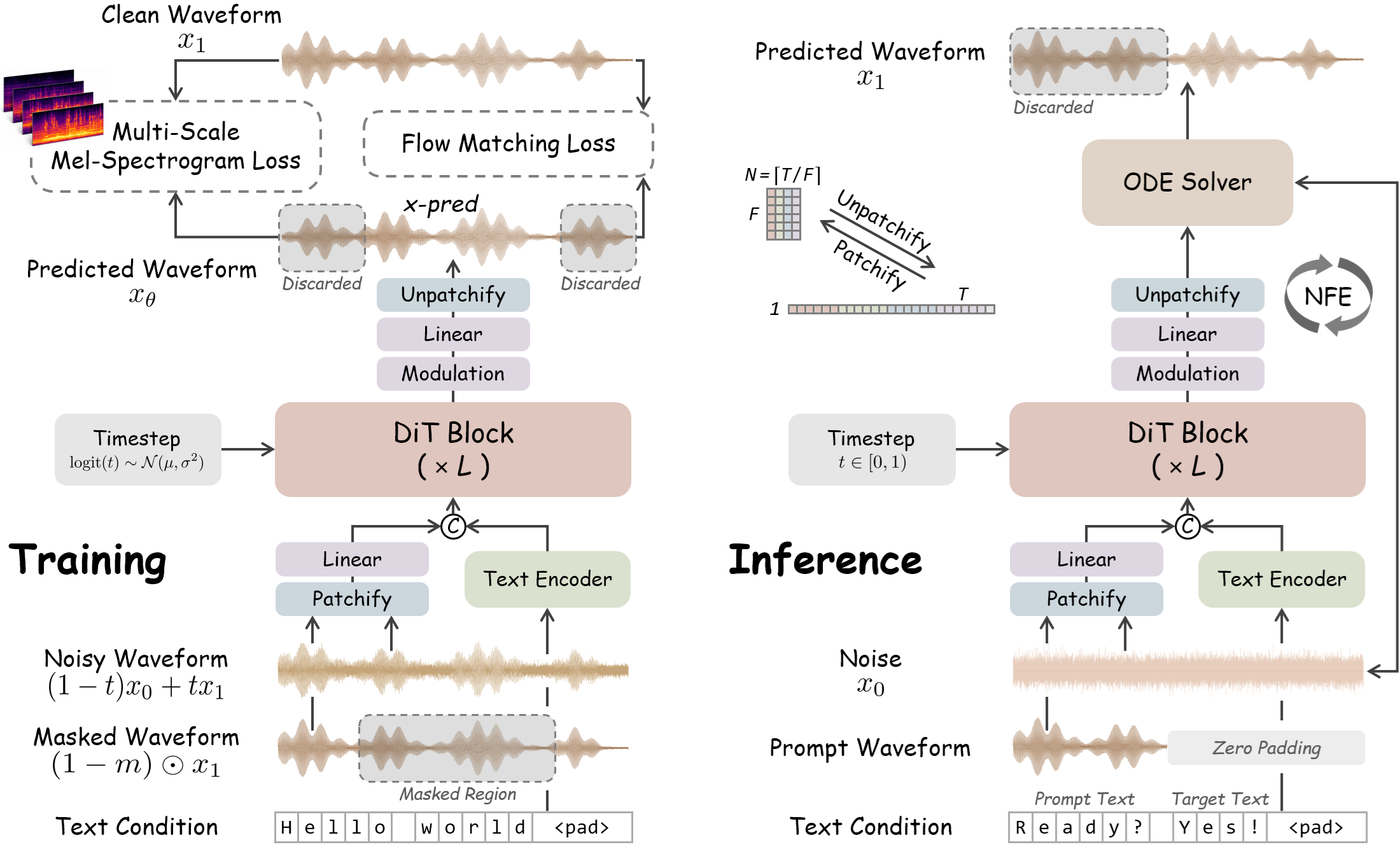}
    \caption{Illustration of WavTTS training (left) and inference (right).}
    \label{fig:wavtts}
    \vspace{-3mm}
\end{figure}

As illustrated in Figure~\ref{fig:wavtts}, WavTTS is an NAR zero-shot TTS model that directly generates raw waveforms. 
This section first introduces the waveform diffusion modeling framework of WavTTS in Section~\ref{sec:modeling_overview}, then describes the multi-scale mel-spectrogram auxiliary supervision in Section~\ref{sec:mel_loss}, and finally presents our noise-aware schedule design for both training and inference in Section~\ref{sec:noise_schedule}.

\subsection{Raw Waveform Diffusion Modeling}
\label{sec:modeling_overview}

Following recent NAR TTS frameworks~\citep{E2-TTS, F5-TTS, LongCat-AudioDiT}, WavTTS models waveform generation under the flow matching (FM) paradigm~\citep{FlowMatching} from the perspective of ordinary differential equations (ODEs). 
Under the rectified flow formulation with linear interpolation~\citep{FlowMatching, liu2022flow}, given a clean speech waveform $x_1 \sim p(x_1)$ from the data distribution and a Gaussian noise sample $x_0 \sim p(x_0)=\mathcal{N}(0,I)$, the intermediate noisy waveform at timestep $t$ is defined as $x_t = (1-t)x_0 + t x_1$, where $t \in [0,1]$ is sampled to learn the transport process from noise to data.
Taking the derivative of $x_t$ with respect to $t$ yields the ground-truth velocity field $v_t = x_1 - x_0$. 
The original FM objective trains a neural network $v_\theta(x_t,t)$ to directly regress this velocity field:

\begin{equation}
\label{eq:fm_v}
\mathcal{L}_{\mathrm{FM}}
=
\mathbb{E}_{t,x_0,x_1}
\left[
\left\|
v_\theta(x_t,t) - v_t
\right\|_2^2
\right].
\end{equation}

However, directly predicting $x_1 - x_0$ requires the model to fit a target containing the stochastic noise component $x_0$, which is particularly challenging in the high-dimensional and complex waveform space. 
This issue becomes more pronounced in silent segments or low-energy frequency regions, where noise-dominated targets may lead to unstable optimization~\citep{Flow2GAN}. 
To mitigate this problem, inspired by JiT~\citep{JiT}, we reformulate the prediction target as directly estimating the clean waveform, i.e., the network outputs $x_\theta = \mathrm{net}_\theta(x_t,t)$. 
Under this formulation, the predicted and ground-truth velocity fields can be rewritten as $v_\theta = \frac{x_\theta - x_t}{1-t}$ and $v_t = \frac{x_1 - x_t}{1-t} = x_1 - x_0$, respectively.
Substituting them into Eq.~\eqref{eq:fm_v}, the original FM objective can be equivalently transformed into the following $x$-prediction objective:

\begin{equation}
\label{eq:fm_x}
\mathcal{L}_{\mathrm{FM}}
=
\mathbb{E}_{t,x_0,x_1}
\left[
\left\|
\frac{x_\theta - x_1}{1-t}
\right\|_2^2
\right].
\end{equation}

To enable zero-shot voice cloning, we employ the text-conditioned speech-infilling task~\citep{Voicebox}, where the model predicts the masked speech segment given the surrounding audio context and the full text transcript. Let $x_1 \in \mathbb{R}^{T}$ denote the clean waveform and $y$ denote the corresponding text transcript. 
As shown in Figure~\ref{fig:wavtts}, we apply a contiguous span mask $m \in \{0,1\}^{T}$, utilizing $x_{\mathrm{ctx}} = (1-m)\odot x_1$ as the audio prompt. Meanwhile, the noisy waveform is obtained by linear interpolation as $x_t=(1-t)x_0+t x_1$.
For efficient temporal modeling, we patchify the 1D raw waveform into non-overlapping blocks of length $F$, yielding a representation in $\mathbb{R}^{N \times F}$, where $N=\lceil T/F\rceil$~\citep{Wave-tacotron, TS3-Codec, MagiCodec}. 
The patchified noisy waveform and audio prompt are then embedded via two-layer linear projections.
For the text condition $y$, represented by bilingual pinyin and alphabet tokens, we pad the text sequence with filler tokens to match the length of the audio patches, enabling implicit text-audio alignment~\citep{E2-TTS, F5-TTS}. 
The text sequence is encoded by ConvNeXt V2 blocks~\citep{ConvNeXt-V2} and concatenated with the audio embeddings along the feature dimension as input to the flow matching network.
Finally, the network output is projected by a linear layer and reshaped, i.e., unpatchified, to recover the predicted waveform $x_\theta$.
Accordingly, the $x$-prediction FM objective in Eq.~\eqref{eq:fm_x} can be reformulated as:

\begin{equation}
\label{eq:fm_x_tts}
\mathcal{L}_{\mathrm{FM}}
=
\mathbb{E}_{t,x_0,x_1}
\left[
\left\|
\frac{
\left(x_\theta(x_t,t,x_{\mathrm{ctx}},y)-x_1\right)\odot m
}{1-t}
\right\|_2^2
\right].
\end{equation}

WavTTS employs DiT~\citep{DiT} as the flow matching backbone, with the sampled timestep $t$ injected via adaLN-Zero conditioning. 
RMSNorm~\citep{rmsnorm} and RoPE~\citep{rope} are applied across all Transformer layers. 
To enable classifier-free guidance (CFG)~\citep{cfg} during inference, we jointly drop the text transcript and audio prompt with a probability of $0.1$ during training, allowing the model to learn an unconditional distribution. 
To avoid division-by-zero instability in Eq.~\eqref{eq:fm_x_tts}, we clip $t$ to a maximum value of $0.98$ when computing the loss.

For inference, we employ the Euler method as the ODE solver, which iteratively transports a randomly sampled initial noise $x_0$ toward the target data distribution $p(x_1)$ via first-order numerical integration.
Given a predefined discrete time schedule $0=t_0<\cdots<t_i<\cdots<t_K=1$, where $K$ denotes the number of function evaluations (NFE), each update step is computed as:

\begin{equation}
\label {eq:euler_update}
x_{t_{i+1}} = x_{t_i} + (t_{i+1} - t_i) v_\theta (x_{t_i}, t_i, x_{\mathrm{ctx}}, y) = x_{t_i} + (t_{i+1} - t_i) \frac {x_\theta (x_{t_i}, t_i, x_{\mathrm{ctx}}, y) - x_{t_i}}{1 - t_i}.
\end{equation}

In practice, we apply CFG by linearly extrapolating the conditional and unconditional $x$-predictions:

\begin{equation}
\label{eq:cfg}
\tilde{x}_\theta(x_{t}, t, x_{\mathrm{ctx}}, y, \alpha) = x_\theta(x_{t}, t, x_{\mathrm{ctx}}, y) + \alpha \Big ( x_\theta(x_{t}, t, x_{\mathrm{ctx}}, y) - x_\theta(x_{t}, t, \emptyset, \emptyset) \Big),
\end{equation}

where $\alpha$ is the CFG scale, and $\emptyset$ denotes that the corresponding condition is replaced with zero padding.
For zero-shot generation, we concatenate the transcript of the reference audio $y_{\mathrm{ref}}$ and the target text $y_{\mathrm{gen}}$ as the text condition. 
The target speech duration is estimated according to the character-length ratio between $y_{\mathrm{ref}}$ and $y_{\mathrm{gen}}$, and is used to determine the length of the masked region after the prompt waveform. 
The model then generates the target speech within this masked region, producing the final waveform prediction.

\subsection{Multi-Scale Mel-Spectrogram Loss}
\label{sec:mel_loss}
Directly modeling high-dimensional waveforms suffers from substantial information redundancy. 
Relying solely on the time-domain FM objective may force the model to fit perceptually insignificant sample-level variations, thereby hindering efficient optimization. 
Prior studies on vocoders~\citep{HiFi-GAN, ParallelWavegan} and neural audio codecs~\citep{DAC, SoundStream, Encodec} have shown that frequency-domain objectives, such as STFT or mel-spectrogram losses~\citep{stftloss}, can effectively improve the perceptual quality of synthesized audio and better align with human auditory perception.

Motivated by these observations, we introduce a multi-scale mel-spectrogram loss as auxiliary perceptual supervision.
Benefiting from the $x$-prediction objective, we can directly compute the log-mel spectrogram distance between the predicted waveform $x_\theta$ and the ground-truth $x_1$ across multiple spectral resolutions. 
This supervision encourages the model to capture both local acoustic details and global spectral structures. 
Consistent with the FM objective, we apply the mel-spectrogram loss only to the masked target speech regions, leading to the following formulation:

\begin{equation}
\label{eq:ms_mel_loss}
\mathcal{L}_{\mathrm{mel}} = \mathbb{E}_{t, x_0, x_1} \left[ \sum_{s \in \mathcal{S}} \frac{\left\| m^{(s)} \odot \left( \Phi_s(x_1) - \Phi_s(x_\theta) \right) \right\|_1}{\| m^{(s)} \|_1} \right],
\end{equation}

where $\mathcal{S}$ denotes the set of mel-spectrogram configurations, $\Phi_s(\cdot)$ extracts the log-mel spectrogram at scale $s$, and $m^{(s)}$ is the span mask aligned with the corresponding temporal resolution. The overall training objective of WavTTS is defined as:

\vspace{-2mm}
\begin{equation}
\label{eq:total_loss}
\mathcal{L}
=
\mathcal{L}_{\mathrm{FM}}
+
\lambda_{\mathrm{mel}}
\mathcal{L}_{\mathrm{mel}},
\end{equation}

where $\lambda_{\mathrm{mel}}$ controls the strength of the perceptual guidance. 
Empirically, we find that this auxiliary loss substantially accelerates model convergence and improves speech naturalness,  without requiring any pre-trained acoustic representation models.

\subsection{Schedule Design for Raw Waveform Flow Matching}
\label{sec:noise_schedule}

Prior studies in computer vision~\citep{simplediffusion, SID2, noise_schedule} have shown that proper noise scheduling is crucial for high-dimensional pixel-space diffusion. We empirically find that this principle is equally important for waveform-space diffusion.
Accordingly, we introduce two noise-aware strategies: \textit{Signal-Noise Variance Alignment} for scale matching (Section~\ref{sec:snr_match}) and \textit{Noise-Shifted Temporal Scheduling} for trajectory adjustment (Section~\ref{sec:noise_shift}).

\subsubsection{Signal-Noise Variance Alignment}
\label{sec:snr_match}

\begin{wrapfigure}{r}{0.45\textwidth}
    \centering
    \vspace{-3mm}
    \includegraphics[width=0.45\textwidth]{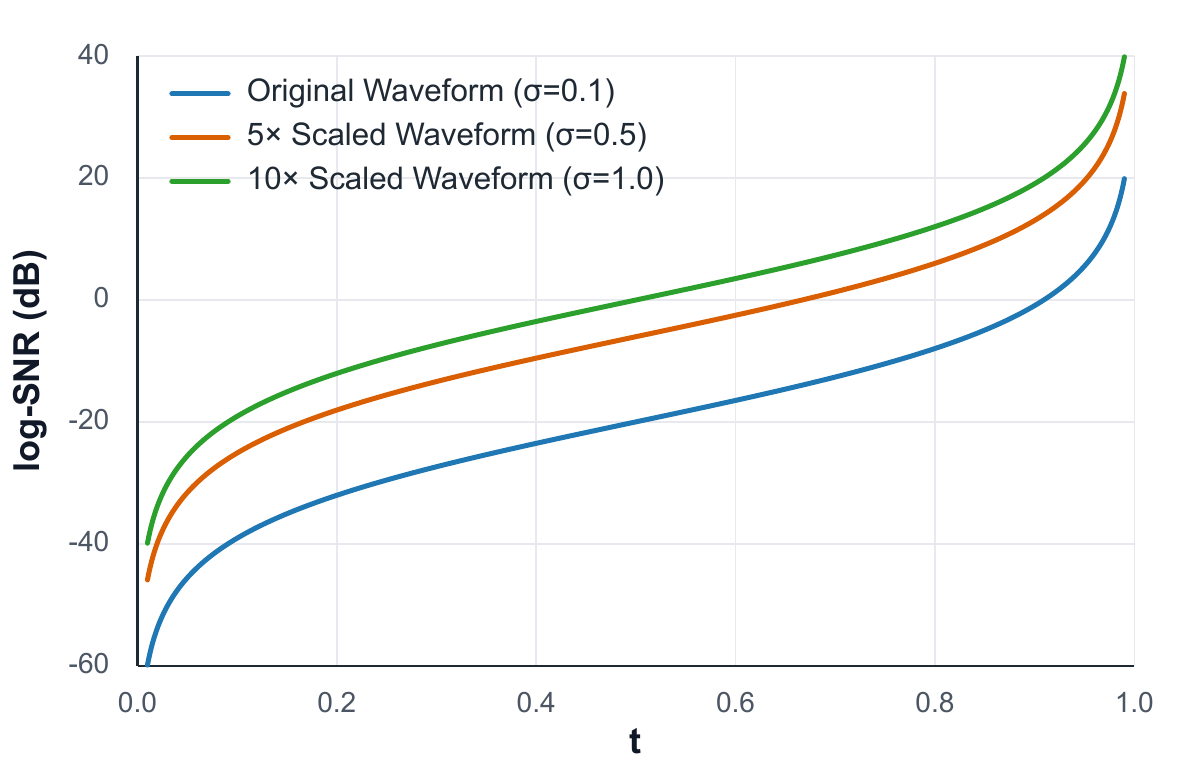}
    \vspace{-5mm}
    \caption{Log-SNR curves under different waveform scaling factors, assuming $\sigma_{x_1}=0.1$.}
    \label{fig:logsnr_flow_matching}
\end{wrapfigure}

Standard rectified flow formulations implicitly assume that the target data distribution and the noise prior have comparable scale. 
However, direct waveform modeling inherently violates this assumption. 
While raw audio is typically bounded within $[-1, 1]$, its empirical standard deviation is much smaller due to the prevalence of silent intervals and low-energy speech regions. 
For example, our statistical analysis shows that the waveform standard deviation is only $\sim 0.12$ on Emilia~\citep{Emilia} and $\sim 0.07$ on LibriTTS~\citep{LibriTTS}. 
This severe scale mismatch between the waveform distribution and the unit Gaussian prior ($\sigma_{x_0} = 1$) leads to a suboptimal signal-to-noise ratio (SNR) trajectory. 
Specifically, under the linear interpolation path $x_t=(1-t)x_0+t x_1$, the Log-SNR can be mathematically decomposed as:

\begin{equation}
\label{eq:log_snr}
\mathrm{Log\text{-}SNR}(t) = 10 \log_{10} \left( \frac{t^2 \sigma_{x_1}^2}{(1-t)^2 \sigma_{x_0}^2} \right) = 20 \log_{10} \left( \frac{t}{1-t} \right) + 20 \log_{10} \left( \frac{\sigma_{x_1}}{\sigma_{x_0}} \right).
\end{equation}

As illustrated in Figure~\ref{fig:logsnr_flow_matching} and Eq.~\eqref{eq:log_snr}, assuming $\sigma_{x_1} = 0.1$ and $\sigma_{x_0}=1$, the actual log-SNR trajectory is shifted downward by $20$~dB compared to the variance-aligned case where $\sigma_{x_1}=\sigma_{x_0}$.
As a result, the model is forced to operate in extremely low-SNR regimes for most training timesteps, making it difficult to recover fine-grained waveform structures from an overwhelming noise background.
Moreover, during inference, transporting unit Gaussian noise to a low-variance waveform distribution is prone to amplifying minor prediction errors, leading to perceptible background noise and unstable artifacts.

To address this issue, we introduce \textit{Signal-Noise Variance Alignment}. 
Before training, we apply a constant scaling factor $k$ to the target waveform, i.e., $x'_1 = k \cdot x_1$, such that $\sigma_{x'_1} \approx 1$. 
We then replace $x_1$ in Eq.~\eqref{eq:fm_x_tts} with the scaled waveform $x'_1$ as the FM prediction target. 
This simple operation removes the negative offset term in Eq.~\eqref{eq:log_snr} without changing the underlying structure of the waveform manifold, providing a smoother and more balanced signal-to-noise trajectory over $t\in(0,1)$. 
Notably, to prevent energy scaling from distorting the perceptual objective, the mel-spectrogram loss is computed strictly at the original waveform scale by comparing $x_1$ with $x_\theta/k$.
During inference, once the model predicts the scaled waveform $x'_1$, we simply apply the inverse scaling factor $1/k$ to recover the audio to its original amplitude range.

\begin{figure}[htbp]
    \centering
    \begin{subfigure}[t]{0.48\linewidth}
        \centering
        \includegraphics[width=\linewidth]{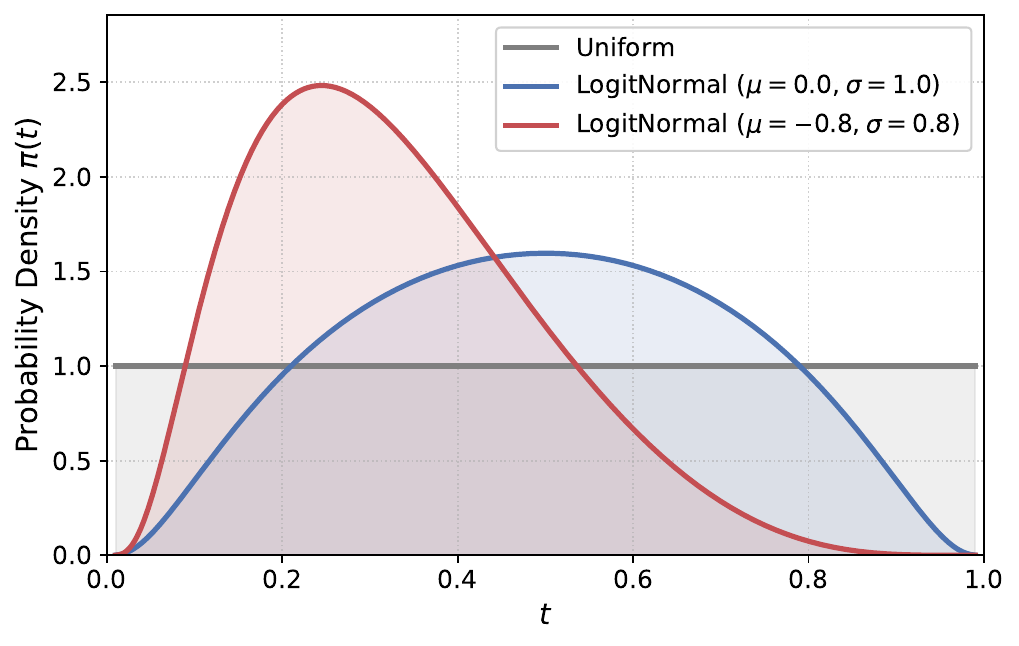}
        \caption{Timestep sampling density $\pi(t)$.}
        \label{fig:time_density}
    \end{subfigure}
    \hfill
    \begin{subfigure}[t]{0.48\linewidth}
        \centering
        \includegraphics[width=\linewidth]{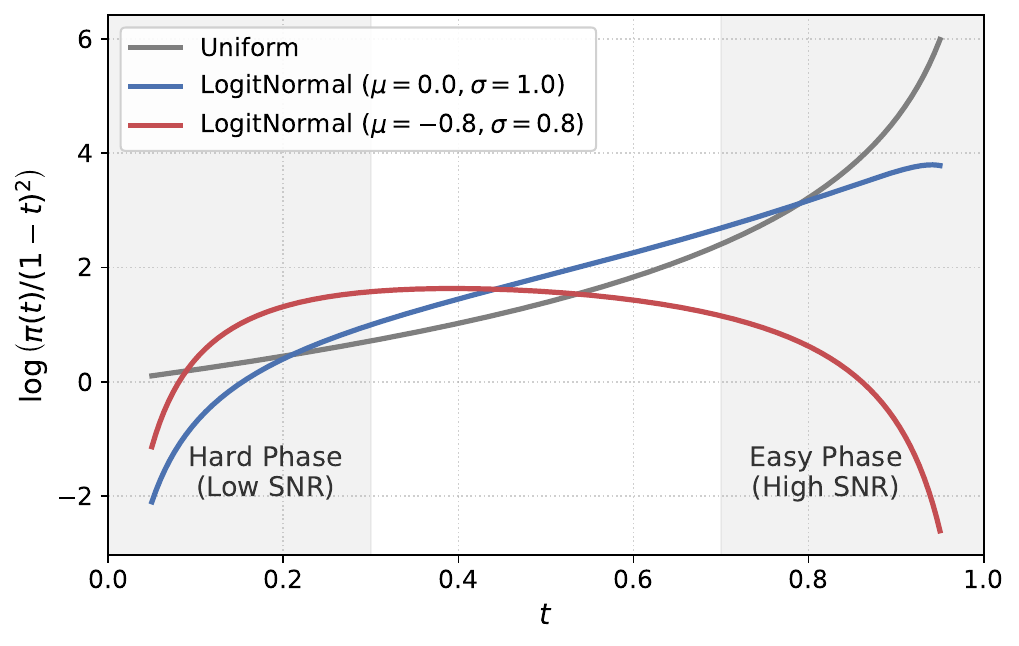}
        \caption{Implicit loss weight $\pi(t)/(1-t)^2$ in log scale.}
        \label{fig:time_weight}
    \end{subfigure}
    \caption{
    Illustration of timestep sampling densities and their corresponding implicit loss weights under the $x$-prediction objective. 
    The noise-shifted schedule emphasizes low-SNR regions while reducing the excessive loss weight near $t\rightarrow1$.
    }
    \label{fig:time_schedule}
\end{figure}
\vspace{-3mm}

\subsubsection{Noise-Shifted Temporal Scheduling}
\label{sec:noise_shift}

During training, existing flow-matching-based TTS models~\citep{F5-TTS, ZipVoice} typically adopt uniform timestep sampling, i.e., $t\sim\mathcal{U}(0,1)$.
However, under the $x$-prediction objective, predicting the raw waveform $x'_1$ from the interpolated state $x_t=(1-t)x_0+t x'_1$ leads to highly imbalanced modeling difficulty across timesteps.
When $t\rightarrow0$, the input is dominated by Gaussian noise, forcing the model to reconstruct the waveform under extremely low-SNR conditions. 
In contrast, when $t\rightarrow1$, the input is already close to the target, and the model only needs to remove minor residual noise. 
To mitigate this issue, following practices in image generation~\citep{ScalingRectifiedFlow, JiT}, we sample training timesteps from a logit-normal distribution.
Formally, we draw $u\sim\mathcal{N}(\mu,\sigma^2)$ and set $t=\mathrm{sigmoid}(u)$, yielding $t\sim\mathrm{LogitNormal}(\mu,\sigma^2)$.
In practice, we set $\mu<0$ to shift the sampling density $\pi(t)$ toward high-noise regions, as shown in Figure~\ref{fig:time_density}.
From the perspective of the $x$-prediction objective, non-uniform timestep sampling can be equivalently interpreted as implicit loss re-weighting~\citep{ScalingRectifiedFlow} under uniform sampling:

\vspace{-3mm}
\begin{equation}
\label{eq:fm_x_weight}
\mathcal{L}_{\mathrm{FM}}
=
\mathbb{E}_{t \sim \pi(t),\,x_0,x'_1}
\left[
\frac{1}{(1-t)^2}\left\|x_\theta - x'_1\right\|_2^2
\right]
= 
\mathbb{E}_{t \sim \mathcal{U}(0,1),\,x_0,x'_1}
\left[
\frac{\pi(t)}{(1-t)^2}\left\|x_\theta - x'_1\right\|_2^2
\right].
\end{equation}

As illustrated in Figure~\ref{fig:time_weight}, noise-shifted scheduling amplifies the expected loss weights in early high-noise regions while suppressing those in later near-clean regions.
This encourages the model to allocate more capacity to the challenging stage of coarse waveform structure formation, while reducing redundant optimization on easier late-stage denoising. 
Notably, the logit-normal density naturally decays near both boundaries, i.e., $t\rightarrow0$ and $t\rightarrow1$. 
This boundary decay avoids over-penalizing nearly pure-noise inputs while also mitigating excessive optimization on near-clean states, encouraging the model to focus on structurally informative low-SNR timesteps.

During inference, shifting the sampling schedule toward high-noise regions, as opposed to uniform sampling, has proven effective for improving speech synthesis quality~\citep{F5-TTS, AcceleratingF5}, a trend that we also observe in direct waveform modeling.
From the perspective of ODE integration, early steps in noisy regions are particularly critical: truncation errors accumulated in this stage can propagate throughout the trajectory and substantially degrade the global waveform structure. 
However, existing inference schedules are not always optimal for our setting.
For example, Sway Sampling~\citep{F5-TTS} provides only a limited degree of timestep shifting, which we find insufficient for high-dimensional raw waveform generation.
To address this limitation, we propose \textit{PolyShift}, a composite noise-shifted inference schedule that combines a polynomial transformation~\citep{FlowTS} with a time-shift function~\citep{ScalingRectifiedFlow}. 
This design enables more flexible control over the inference trajectory and allows denser integration in challenging high-noise regions.
Given a uniformly spaced sequence $\tau\in[0,1]$, the actual inference timestep $t$ is defined as:

\vspace{-3mm}
\begin{equation}
\label{eq:infer_shift}
t = \frac{\tau^p}{\tau^p + s (1 - \tau^p)},
\end{equation}
\vspace{-3mm}

where $p$ is the power factor and $s$ is the shift factor. 
By setting $p>1$ and $s>1$, PolyShift sampling flexibly allocates more integration steps to challenging high-noise regions. 
This reduces early-stage numerical errors and suppresses generation artifacts, ultimately leading to higher-fidelity waveform synthesis.
We provide further comparison and analysis of different inference timestep sampling strategies in Appendix~\ref{app:infer_schedule}.

%% file: sections/experiments_setup.tex
\section{Experimental Setup}
\label{sec:exp_setup}

\textbf{Datasets.}
We train WavTTS on the open-source Emilia dataset~\citep{Emilia}, which contains approximately 95K hours of English and Chinese speech. 
For zero-shot TTS evaluation, we use the Seed-TTS \textit{test-en} set~\citep{Seed-TTS}, which consists of 1,088 samples from Common Voice~\citep{CommonVoice}, and the Seed-TTS \textit{test-zh} set, which contains 2,020 samples from DiDiSpeech~\citep{Didispeech}. 
For standard TTS evaluation against prior end-to-end speech generation models, we use 682 in-domain English test samples from LJSpeech~\citep{ljspeech} and the LibriSpeech-PC~\citep{LibriSpeech-PC} \textit{test-clean} subset, which contains 1,127 English samples, following the evaluation split of F5-TTS~\citep{F5-TTS}.

\textbf{Model Setup.}
We train WavTTS for 1.2M steps on 8 NVIDIA A100 80GB GPUs, with a batch size of 153,600 audio patch frames, corresponding to approximately $0.43$ hours of audio. 
We use the AdamW optimizer~\citep{adamw} with a peak learning rate of $7.5\times10^{-5}$, which is linearly warmed up for 20K updates and then kept constant.
All audio is resampled to 16 kHz, and the waveform patch size is set to $F=160$, resulting in a patchified sequence rate of 100 Hz. 
We set $\lambda_{\mathrm{mel}}=0.05$ to balance the FM objective and mel-spectrogram supervision.
For noise-aware training, we use a scaling factor of $k=9$, since Emilia's empirical waveform standard deviation is approximately $0.12$.
Following prior practice~\citep{JiT}, we adopt logit-normal timestep sampling with $\mu=-0.8$ and $\sigma=0.8$.
During inference, we use 50 NFEs, a CFG scale of $\alpha=3$, and the PolyShift inference schedule with $p=2$ and $s=3$.
More model configurations and implementation details are provided in Appendix~\ref{app:implement_details}.

\textbf{Baselines.}
For zero-shot TTS evaluation, we primarily compare WavTTS with state-of-the-art NAR models, including mel-spectrogram-based systems such as E2-TTS~\citep{E2-TTS}, F5-TTS~\citep{F5-TTS}, and ZipVoice~\citep{ZipVoice}, as well as latent-based systems such as MaskGCT~\citep{MaskGCT} and LongCat-AudioDiT~\citep{LongCat-AudioDiT}. 
We also report the performance of representative AR models for reference, including CosyVoice~\citep{CosyVoice}, CosyVoice 2~\citep{CosyVoice2}, Llasa~\citep{Llasa}, and Spark-TTS~\citep{Spark-TTS}. 
For standard TTS evaluation against prior end-to-end waveform generation models, we use publicly available implementations, including WaveGrad 2~\citep{WaveGrad2}, VITS~\citep{VITS}, and JETS~\citep{JETS}. 
For VITS, we evaluate two variants trained on LJSpeech and VCTK~\citep{VCTK}, denoted as VITS$_{\mathrm{LJ}}$ and VITS$_{\mathrm{VCTK}}$, respectively.

\textbf{Evaluation Metrics.}
We adopt three reproducible model-based metrics for objective evaluation. 
Intelligibility is measured by word error rate (WER) using ASR models, with Whisper-large-v3~\citep{whisper} for English and Paraformer-zh~\citep{Paraformer} for Chinese. 
Speaker similarity is evaluated by SIM-o, where a WavLM-based speaker verification model~\citep{sim-o} is used to extract speaker representations from the generated speech and the reference prompt, followed by cosine similarity computation. 
Naturalness is assessed by UTMOS~\citep{UTMOS}, which predicts the mean opinion score (MOS) of synthesized speech.

%% file: sections/main_results.tex
\section{Experimental Results}
\vspace{-1mm}

\begin{table*}[t]
\centering
\caption{
Zero-shot TTS results on the Seed-TTS benchmark.
Best results are highlighted in \textbf{bold}, and second-best are \underline{underlined}.
``Multi.'' denotes multilingual training data.
$^\dagger$Results are taken from the original papers.
}
\resizebox{1\linewidth}{!}{
\begin{tabular}{l c c ccc ccc}
\toprule
\multirow{2}{*}{\textbf{Model}} & \multirow{2}{*}{\textbf{Params}} & \multirow{2}{*}{\textbf{Data (hrs)}} & \multicolumn{3}{c}{\textbf{Seed-TTS test-en}} & \multicolumn{3}{c}{\textbf{Seed-TTS test-zh}} \\
\cmidrule(lr){4-6} \cmidrule(lr){7-9}
& & & \textbf{WER(\%)} $\downarrow$ & \textbf{SIM-o} $\uparrow$ & \textbf{UTMOS} $\uparrow$ & \textbf{CER(\%)} $\downarrow$ & \textbf{SIM-o} $\uparrow$ & \textbf{UTMOS} $\uparrow$ \\
\midrule
Ground Truth & -- & -- & 1.79 & 0.73 & 3.53 & 1.25 & 0.75 & 2.78 \\
\shline
\rowcolor{gray!15}\multicolumn{9}{l}{\textbf{\textit{AR Models$^\dagger$}}} \\
CosyVoice & 416M & 170K Multi. & 4.29 & 0.61 & -- & 3.63 & 0.72 & -- \\
CosyVoice 2 & 618M & 167K Multi. & 2.57 & 0.65 & -- & 1.45 & 0.75 & --\\
Llasa-1B & 1370M & 250K Multi. & 3.22 & 0.57 & -- & 1.89 & 0.67 & -- \\
Spark-TTS & 507M & 102K Multi. & 1.98 & 0.58 & -- & \underline{1.20} & 0.67 & -- \\
\shline
\rowcolor{gray!15}\multicolumn{9}{l}{\textbf{\textit{NAR Latent/Mel-Spectrogram Models}}} \\
MaskGCT & 1048M & 100K Emilia & 2.36 & \underline{0.71} & 3.57 & 2.48 & \underline{0.77} & 2.64 \\
E2-TTS & 333M & 100K Emilia & 2.21 & \underline{0.71} & 3.20 & 1.97 & 0.73 & 2.27\\
F5-TTS & 336M & 100K Emilia & 1.65 & 0.66 & 3.73 & 1.55 & 0.75 & 2.94 \\
ZipVoice & 123M & 100K Emilia & \underline{1.60} & 0.70 & \underline{3.83} & 1.40 & 0.75 & \underline{3.15} \\
LongCat-AudioDiT & 1420M & 100K Multi. & 1.94 & \textbf{0.76} & 3.80 & \textbf{1.10} & \textbf{0.81} & \textbf{3.16} \\
\shline
\rowcolor{gray!15}\multicolumn{9}{l}{\textbf{\textit{NAR Waveform Space Models}}} \\
WavTTS & 673M & 100K Emilia & \textbf{1.50} & 0.65 & \textbf{3.92} & 1.59 & 0.73 & 3.08 \\ 
\bottomrule
\end{tabular}
}
\label{tab:zero_shot_tts_results}
\vspace{-3mm}
\end{table*}

\subsection{Overall Performance}
\vspace{-1mm}
\subsubsection{Zero-Shot TTS Evaluation}

As shown in Table~\ref{tab:zero_shot_tts_results}, WavTTS demonstrates that high-quality zero-shot TTS can be achieved without relying on pre-trained neural codecs, vocoders, or autoencoders.
In particular, it exhibits strong intelligibility and objective naturalness. 
On the English test set, it achieves the best WER of 1.50\% and the highest UTMOS score of 3.92 among all evaluated baselines, while also maintaining competitive performance on the Chinese test set. 
In terms of speaker similarity, WavTTS still lags behind some highly optimized latent-space models.
We hypothesize that this is because waveform-space generation operates on high-dimensional, uncompressed time-domain signals that contain rich, fine-grained acoustic variations (e.g., phase and ambient details). This inherent complexity makes it harder for a finite-capacity model to prioritize target-speaker timbre cloning during generation.
Therefore, further model scaling and tailored speaker-oriented alignment strategies may be important for fully exploiting the potential of raw waveform TTS.

Overall, the results validate the feasibility of direct waveform-space generation for zero-shot TTS, establishing WavTTS as a streamlined end-to-end alternative to existing multi-stage TTS pipelines.

\subsubsection{Comparison with End-to-End Speech Generation Models}

\begin{wraptable}{r}{0.52\textwidth}
\centering
\vspace{-3mm}
\caption{Comparison of TTS performance with previous end-to-end speech generation models.}
\label{tab:e2e_tts_comparison}
\resizebox{0.52\textwidth}{!}{
\begin{tabular}{l cccc}
\toprule
\multirow{2}{*}{\textbf{Model}} 
& \multicolumn{2}{c}{\textbf{LJSpeech}} 
& \multicolumn{2}{c}{\textbf{LibriSpeech-PC}} \\
\cmidrule(lr){2-3} \cmidrule(lr){4-5}
& \textbf{WER(\%)} $\downarrow$ & \textbf{UTMOS} $\uparrow$
& \textbf{WER(\%)} $\downarrow$ & \textbf{UTMOS} $\uparrow$ \\
\midrule
Ground Truth & 3.42 & 4.36 & 2.23 & 4.10 \\
\midrule
WaveGrad 2 & 25.19 & 3.24 & 33.77 & 3.05 \\
VITS$_{\mathrm{VCTK}}$ & 9.34 & 4.06 & 10.23 & 4.03 \\
VITS$_{\mathrm{LJ}}$ & 3.72 & 4.37 & 2.23 & \textbf{4.36} \\
JETS & 3.73 & 4.36 & 3.00 & 4.34 \\
\midrule
WavTTS & \textbf{3.43} & \textbf{4.39} & \textbf{2.02} & \textbf{4.36} \\
\bottomrule
\end{tabular}
}
\vspace{-3mm}
\end{wraptable}

Since most prior end-to-end speech generation models do not support zero-shot TTS, we focus our comparison on intelligibility (WER) and objective naturalness (UTMOS).
These baselines are predominantly trained on the single-speaker LJSpeech~\citep{ljspeech} dataset, except for VITS$_{\mathrm{VCTK}}$.
To reduce the influence of speaker timbre differences on the UTMOS metric and ensure a fairer naturalness evaluation, we prompt WavTTS with a fixed audio clip randomly selected from LJSpeech for zero-shot synthesis.

As shown in Table~\ref{tab:e2e_tts_comparison}, WavTTS achieves the lowest WER and the highest UTMOS across both test sets.
Notably, under the zero-shot setting, WavTTS outperforms prior supervised systems on the LJSpeech test set, which closely matches their training distributions, and maintains strong performance on the out-of-domain LibriSpeech-PC dataset.
These results demonstrate the superior intelligibility, naturalness, and robust generalization capabilities of our approach.

It is worth noting that although models such as VITS and JETS are commonly categorized as end-to-end systems, they do not model raw audio directly.
Instead, their core acoustic sequence modeling is performed in intermediate latent spaces with lower temporal resolution, followed by adversarial decoders that upsample these representations into high-dimensional waveforms.
In contrast, WavTTS directly models raw waveforms through patchification while achieving superior TTS performance. 
This demonstrates the potential of native waveform generation as a simpler and more direct paradigm for end-to-end speech synthesis.

%% file: sections/ablation_study.tex
\subsection{Ablation Studies}
We conduct ablation studies on WavTTS to validate the effectiveness of our core design choices.
For a fair and computationally efficient comparison, all variants are trained for 1M steps on 8 NVIDIA A100 80GB GPUs and evaluated on the Seed-TTS \textit{test-en} set with the same inference strategy as in Section~\ref{sec:exp_setup}.

\subsubsection{Training Objectives}

\begin{wraptable}{r}{0.52\textwidth}
\centering
\vspace{-0.8em}
\caption{Ablations on the prediction objective and mel-spectrogram loss weight. Default settings are marked in \colorbox{gray!15}{gray}.}
\label{tab:ablation_objectives}
\resizebox{0.52\textwidth}{!}{
\begin{tabular}{lcccc}
\toprule
\textbf{Prediction Target} & $\bm{\lambda_{\mathrm{mel}}}$ & \textbf{WER (\%) $\downarrow$} & \textbf{SIM-o $\uparrow$} & \textbf{UTMOS $\uparrow$} \\
\midrule
$v$-prediction & 0.05 & 1.67 & 0.61 & \textbf{3.94} \\
\midrule
\multirow{5}{*}{$x$-prediction} 
 & 0    & 1.92 & 0.56 & 3.77  \\
 & 0.01 & 1.77 & 0.60 & 3.87  \\
 & 0.05 & \cellcolor{gray!15}\textbf{1.65} & \cellcolor{gray!15}\textbf{0.65} & \cellcolor{gray!15}3.93 \\
 & 0.2  & 1.74 & 0.60 & 3.89  \\
 & 0.5  & 1.81 & 0.55 & 3.82  \\
\bottomrule
\end{tabular}
}
\vspace{-0.8em}
\end{wraptable}

As shown in Table~\ref{tab:ablation_objectives}, we study the impact of the flow matching prediction target and the weight of the multi-scale mel-spectrogram loss $\lambda_{\mathrm{mel}}$.
Compared with $v$-prediction, $x$-prediction achieves slightly better intelligibility and a clear improvement in speaker similarity, suggesting that directly predicting the clean waveform, rather than the vector field, provides a more effective learning target for raw waveform modeling.
In addition, $x$-prediction naturally aligns with auxiliary mel-spectrogram supervision, as the mel loss can be directly applied to the predicted waveform. By contrast, $v$-prediction requires reconstructing the waveform prediction from the estimated velocity before computing the mel loss.

The ablation on $\lambda_{\mathrm{mel}}$ highlights the critical role of appropriate perceptual supervision.
Without the mel-spectrogram loss ($\lambda_{\mathrm{mel}}=0$), the model degrades consistently across all metrics, indicating that sample-level flow matching alone is insufficient for efficient raw waveform generation. 
The mel loss also substantially accelerates training convergence: at 200k steps, the model trained with mel supervision already reduces WER below 5\%, while the model without mel supervision still fails to generate intelligible speech.
However, excessively large values of $\lambda_{\mathrm{mel}}$ ($\lambda_{\mathrm{mel}}\in\{0.2,0.5\}$) degrade both intelligibility and speaker similarity, suggesting that overly strong perceptual supervision may distract the model from learning the underlying waveform-space flow. 
We therefore set $\lambda_{\mathrm{mel}}=0.05$ as the default configuration.

\subsubsection{Noise Scheduling}

\begin{wraptable}{r}{0.35\textwidth}
\centering
\vspace{-3mm}
\caption{Ablations on the scaling factor $k$. Default settings are marked in \colorbox{gray!15}{gray}.}
\label{tab:ablation_scaling_factor}
\resizebox{0.35\textwidth}{!}{
\begin{tabular}{lccc}
\toprule
$\bm{k}$ & \textbf{WER (\%) $\downarrow$} & \textbf{SIM-o $\uparrow$} & \textbf{UTMOS $\uparrow$} \\
\midrule
1  & 4.18 & 0.32 & 2.40 \\
5  & \textbf{1.51} & 0.59 & 3.81 \\
9  & \cellcolor{gray!15}1.65 & \cellcolor{gray!15}\textbf{0.65} & \cellcolor{gray!15}\textbf{3.93} \\ 
10 & 1.82 & 0.64 & 3.87 \\
\bottomrule
\end{tabular}
}
\vspace{-0.8em}
\end{wraptable}

Table~\ref{tab:ablation_scaling_factor} investigates the impact of the waveform scaling factor $k$ on model performance.
Without amplitude scaling ($k=1$), the model suffers from severe degradation, with SIM-o and UTMOS scores substantially lower than those of the scaled variants. 
This indicates that, without proper scaling, the target waveform remains overwhelmed by Gaussian noise over a large portion of the diffusion trajectory, leading to inefficient learning in the raw waveform space during training.
By employing the proposed \textit{Signal-Noise Variance Alignment} strategy ($k=9$), the model achieves the highest SIM-o and UTMOS scores while maintaining a competitive WER.
Interestingly, a smaller scaling factor ($k=5$) slightly improves intelligibility and yields faster WER convergence.
We hypothesize that a relatively higher noise ratio forces the model to prioritize coarse-grained linguistic structures.
However, this intelligibility gain comes at the expense of speaker similarity and naturalness; subjective listening further reveals audible artifacts such as electronic noise and incomplete denoising.
Overall, aligning the variance between the signal and Gaussian noise provides a better balance between objective metrics and perceptual quality, making it an effective design choice for raw waveform diffusion modeling.

\begin{figure}[htbp]
    \centering
    \includegraphics[width=\linewidth]{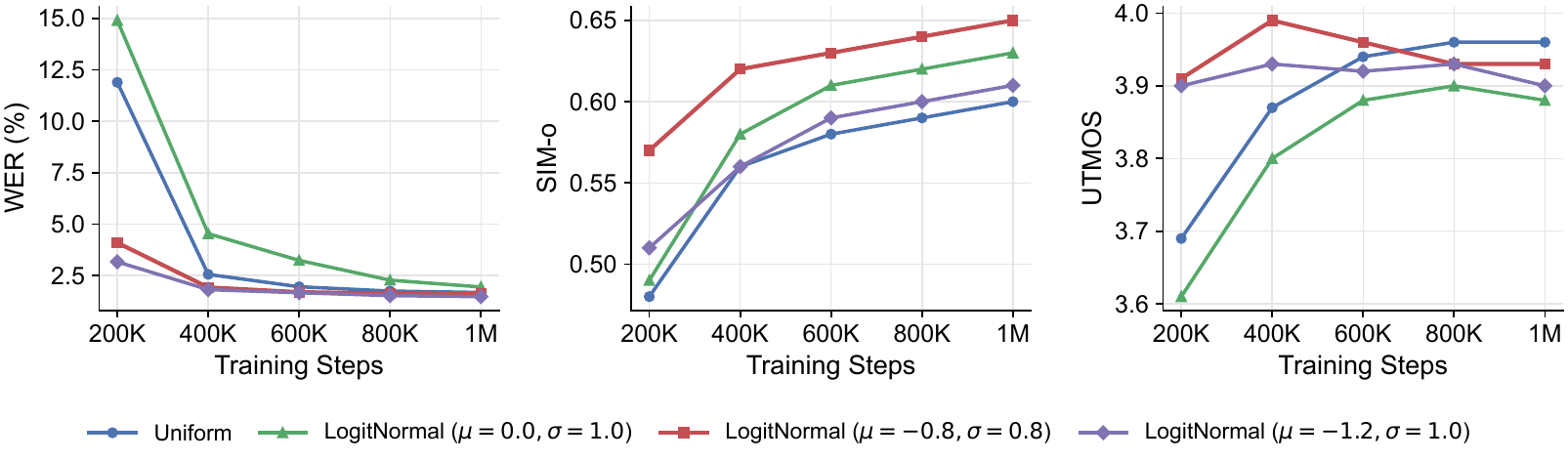}
    \caption{Comparison of zero-shot TTS performance under different training noise schedules.}
    \label{fig:noise_schedule}
\end{figure}

To validate the noise-shifted temporal scheduling introduced in Section~\ref{sec:noise_shift}, we compare different timestep sampling strategies during training. Specifically, we evaluate the three distributions illustrated in Figure~\ref{fig:time_density}, together with a more aggressive noise-shifted variant, $t\sim\mathrm{LogitNormal}(\mu=-1.2,\sigma=1.0)$, which further biases the sampling density toward high-noise regions compared with $\mu=-0.8$.

As shown in Figure~\ref{fig:noise_schedule}, appropriately shifting training timesteps toward high-noise regions accelerates convergence and improves final performance.
In particular, using a logit-normal distribution with $\mu<0$ leads to substantially faster WER convergence than the other strategies, reducing WER below 2\% within 400K steps and achieving a lower final WER.
This suggests that high-noise timesteps are crucial for learning text--speech alignment and coarse linguistic structures in waveform-based flow matching; consequently, increasing their sampling probability enhances synthesis intelligibility.
However, excessively shifting the sampling distribution toward the noise side can compromise the modeling of speaker characteristics and fine-grained acoustic details. 
For example, the aggressive setting, $t\sim\mathrm{LogitNormal}(\mu=-1.2,\sigma=1.0)$, leads to degradation in both SIM-o and UTMOS. 
We therefore adopt $\mu=-0.8$ and $\sigma=0.8$ as the default configuration, which strikes a better balance for zero-shot TTS.
Interestingly, uniform sampling converges more slowly but achieves the highest final UTMOS, possibly because it allocates more training samples to lower-noise regions, which helps the model refine local acoustic details. 
Future work will explore dynamic sampling scheduling that adjusts the timestep distribution over the course of training to further improve overall performance.

\subsubsection{Inference Strategies}

\begin{wraptable}{r}{0.52\textwidth}
\centering
\vspace{-0.8em}
\caption{Comparison of inference timestep schedules. Default settings are marked in \colorbox{gray!15}{gray}.}
\label{tab:ablation_inference_sampling}
\resizebox{0.52\textwidth}{!}{
\begin{tabular}{lccc}
\toprule
\textbf{Inference Schedule} 
& \textbf{WER (\%) $\downarrow$} 
& \textbf{SIM-o $\uparrow$} 
& \textbf{UTMOS $\uparrow$} \\
\midrule
Uniform 
& 1.78 & 0.63 & 3.77 \\
Sway Sampling $(s'=-1.0)$ 
& 1.68 & 0.64 & 3.88 \\
PolyShift $(p=2.0, s=1.0)$ 
& 1.60 & 0.64 & 3.88 \\
PolyShift $(p=2.0, s=3.0)$ 
& \cellcolor{gray!15}1.65 & \cellcolor{gray!15}\textbf{0.65} & \cellcolor{gray!15}\textbf{3.93} \\
PolyShift $(p=2.0, s=5.0)$ 
& \textbf{1.58} & \textbf{0.65} & 3.92 \\
\bottomrule
\end{tabular}
}
\vspace{-0.8em}
\end{wraptable}

We evaluate the model, trained for 1M steps, using various timestep schedules during inference with the NFE set to 50. As shown in Table~\ref{tab:ablation_inference_sampling}, noise-shifted schedules, such as Sway Sampling and our proposed PolyShift, significantly outperform uniform sampling in overall zero-shot TTS performance. This demonstrates that allocating more timesteps to the high-noise regime—namely, the initial phase of the ODE trajectory—crucially enhances final speech synthesis quality. 
Furthermore, applying a more aggressive shift strategy by replacing Sway Sampling with PolyShift yields further improvements in SIM-o and UTMOS. This suggests that flow matching in the raw waveform space necessitates a denser early-stage timestep allocation to formulate a better generative trajectory prototype. 
Notably, while an excessively large shift, such as PolyShift ($p=2.0, s=5.0$), marginally reduces the WER, it introduces more pronounced background noise in subjective listening. 
We attribute this artifact to inadequate fine-grained refinement, caused by a scarcity of timesteps allocated to the later stages. Consequently, we adopt PolyShift ($p=2.0, s=3.0$) as our default inference configuration.

\subsubsection{Scaling Behaviors}

To investigate the scaling behavior of WavTTS with respect to training data and model size, we train WavTTS and a smaller-backbone variant, detailed in Appendix~\ref{app:implement_details}, on two datasets: LibriTTS~\citep{LibriTTS} and Emilia~\citep{Emilia}. 
LibriTTS represents a low-resource setting with approximately 585 hours of English speech, whereas Emilia provides a large-scale 100K-hour multilingual corpus.

\begin{wraptable}{r}{0.52\textwidth}
\centering
\vspace{-0.8em}
\caption{Scaling behavior of WavTTS under different training data scales and model sizes.}
\label{tab:scaling_behaviors}
\resizebox{0.52\textwidth}{!}{
\begin{tabular}{lcccc}
\toprule
\textbf{Training Data} & \textbf{Model Size} & \textbf{WER (\%) $\downarrow$} & \textbf{SIM-o $\uparrow$} & \textbf{UTMOS $\uparrow$} \\
\midrule
\multirow{2}{*}{LibriTTS (585 hrs)} 
 & 340M & 2.16 & 0.35 & \textbf{3.95} \\
 & 673M & 2.12 & 0.31 & 3.94 \\
\midrule
\multirow{2}{*}{Emilia (100K hrs)} 
 & 340M & 1.74 & 0.56 & 3.87 \\
 & 673M & \textbf{1.65} & \textbf{0.65} & 3.93 \\
\bottomrule
\end{tabular}
}
\vspace{-0.8em}
\end{wraptable}

As shown in Table~\ref{tab:scaling_behaviors}, the scale of training data and model capacity both have a substantial impact on final performance.
When trained on LibriTTS, both models exhibit poor zero-shot generalization on the out-of-domain Seed-TTS \textit{test-en} set, yielding SIM-o scores in the 0.3 range.
In contrast, scaling up to the 100K-hour Emilia dataset markedly improves speaker similarity and reduces WER below 2\% for both model sizes.
This suggests that large-scale and diverse data is essential for learning robust speaker characteristics and linguistic content directly in the high-dimensional waveform space.

Increasing the model size from 340M to 673M further improves WER and SIM-o when trained on Emilia, with a particularly clear gain in speaker similarity.
Interestingly, this benefit is not observed under the low-resource LibriTTS setting, where the larger model brings little intelligibility improvement and even slightly degrades speaker similarity. This indicates that model scaling is effective only when supported by sufficient training data. 
Overall, these results show scaling trends consistent with prior large-scale TTS systems~\citep{BASE-TTS, CosyVoice3}, suggesting that sufficient data scale and matched model capacity are both critical for achieving high-quality zero-shot TTS in the raw waveform space.

\begin{figure}[htbp]
    \centering
    \includegraphics[width=\linewidth]{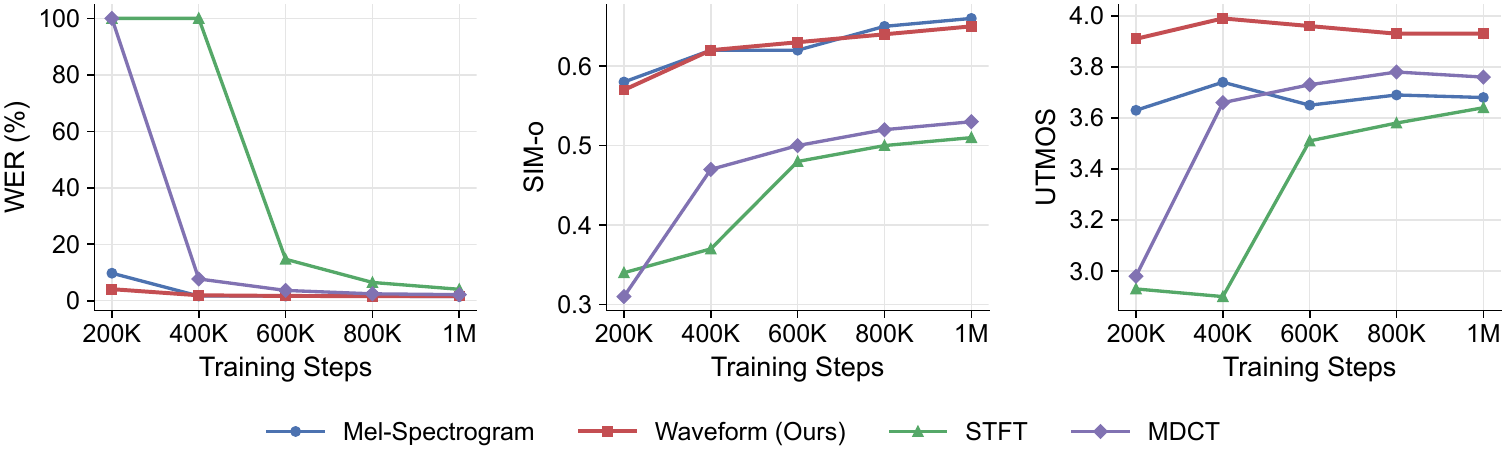}
    \caption{Comparison of zero-shot TTS performance curves using different acoustic representations. Waveform, STFT, and MDCT are lossless (or nearly lossless) representations, while mel-spectrograms are lossy and require an additional pre-trained vocoder~\citep{Vocos} for waveform reconstruction.}
    \label{fig:waveform_representations}
\end{figure}

\subsubsection{Comparison of Acoustic Representations}

Raw waveforms preserve speech signals directly in the time domain, while speech can also be represented by lossless or nearly lossless time-frequency transforms, such as short-time Fourier transform (STFT) and modified discrete cosine transform (MDCT) coefficients, which support waveform reconstruction through corresponding inverse transforms. 
To examine whether frequency-domain modeling offers advantages over direct waveform modeling, we compare these representations under the same flow matching framework.
For reference, we additionally include mel-spectrograms as a standard lossy acoustic representation. 
Details of STFT- and MDCT-based diffusion modeling are provided in Appendix~\ref{app:frequency_domain_diffusion}.

Figure~\ref{fig:waveform_representations} compares zero-shot TTS performance across different acoustic representations.
Overall, both raw waveform and mel-spectrogram models converge efficiently, achieving under 10\% WER and over 0.5 SIM-o at 200K training steps.
Notably, waveform modeling outperforms mel-spectrograms with faster early intelligibility convergence (4.10\% vs. 9.76\% WER at 200K steps) and better objective naturalness (3.93 vs. 3.68 UTMOS at 1M steps), while maintaining comparable speaker similarity.
These results suggest that WavTTS can effectively model high-dimensional waveform structures and benefit from the lossless nature of raw waveforms for natural speech synthesis.

In contrast, STFT and MDCT representations exhibit significantly slower convergence and inferior final zero-shot TTS performance. The MDCT-based model begins to generate recognizable speech after 400K steps, while the STFT-based model requires around 600K steps.
We hypothesize that this stems from the inherent complexity of these time-frequency representations. STFT requires modeling complex-valued coefficients with coupled magnitude-phase structures, while MDCT exhibits an unfavorable feature distribution for flow matching. 
For example, on sampled Emilia data, MDCT features have a standard deviation of only about 0.005 but a maximum value of 0.15, indicating a sharply peaked distribution with extreme outliers that may hinder effective learning.
Consequently, developing flow-matching TTS in such frequency-domain spaces may necessitate more delicate feature preprocessing and architectural modifications.

Overall, this comparison highlights direct waveform modeling as a simple yet effective approach. 
It avoids additional time-frequency transforms and inverse reconstruction, while achieving performance comparable to, or even better than, the widely used lossy mel-spectrogram representation.

%% file: sections/conclusion.tex
\section{Conclusion}
In this paper, we proposed WavTTS, an end-to-end zero-shot TTS framework that directly models speech in the raw waveform space.
By combining flow matching with a Diffusion Transformer backbone and an efficient patchification strategy, WavTTS enables tractable modeling of high-dimensional time-domain signals without relying on neural codecs, vocoders, or autoencoders.
To further improve optimization efficiency and perceptual quality, we adopted an $x$-prediction objective with auxiliary multi-scale mel-spectrogram supervision, together with signal-noise variance alignment and noise-shifted temporal scheduling tailored for waveform-space modeling.
Experimental results demonstrate that WavTTS closely approaches state-of-the-art NAR zero-shot TTS models based on compressed representations, while substantially outperforming previous end-to-end speech generation systems. 
Overall, WavTTS validates the feasibility of a streamlined, native waveform generation paradigm, paving a promising path toward high-quality end-to-end speech synthesis.

%% file: sections/appendix.tex
\section{Implementation Details}
\label{app:implement_details}

We train two model variants with different scales: a large model with 673M parameters as our final configuration, and a smaller model with 340M parameters for scaling ablations. 
Both variants adopt a DiT \citep{DiT} backbone and share the training strategy detailed in Section~\ref{sec:exp_setup}, differing only in size. The large model consists of 28 Transformer layers with a hidden dimension of 1152 and an FFN expansion ratio of 4, while the smaller model uses 22 layers with a hidden dimension of 1024 and an FFN expansion ratio of 2. Both models use 16 attention heads, and the dropout rate in Transformer layers is set to 0.
The text encoder consists of four ConvNeXt V2 blocks \citep{ConvNeXt-V2} with an embedding dimension of 512 and an FFN expansion ratio of 2. 
For the patchified waveforms, we apply a two-layer linear projection without activations: a bias-free layer mapping to a 768-dimensional intermediate representation, followed by a biased layer projecting to 1024 dimensions.
During infilling-task training \citep{Voicebox}, a continuous segment covering 70\%--100\% of the audio prompt is randomly masked.

We adopt the multi-scale mel-spectrogram loss from DAC \citep{DAC} as an additional perceptual supervision during training. Specifically, we compute mel-spectrograms at seven different time--frequency resolutions with window sizes $[32, 64, 128, 256, 512, 1024, 2048]$, where the hop size at each scale is set to one quarter of the corresponding window length. The number of mel bins for each scale is configured as $[5, 10, 20, 40, 80, 160, 320]$. All mel transforms are computed from magnitude spectrograms ($\mathrm{power}=1.0$) with reflective padding and centered STFT computation. For each scale, we compute the $L_1$ distance between the log-mel spectrograms of the predicted waveform and the target waveform over the masked regions, as described in Section~\ref{sec:mel_loss}.

\section{Comparison of Inference Timestep Schedules}
\label{app:infer_schedule}

\begin{wrapfigure}{r}{0.42\textwidth}
    \centering
    \vspace{-2mm}
    \includegraphics[width=0.42\textwidth]{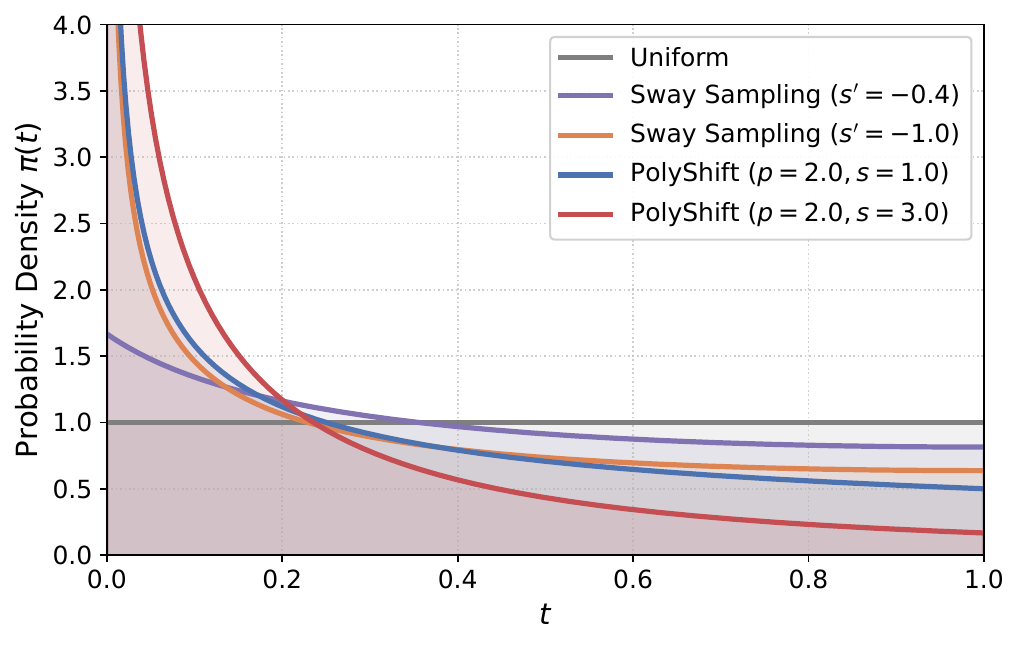}
    \caption{Timestep sampling densities under different inference schedules.}
    \label{fig:infer_sampling_density}
    \vspace{-2mm}
\end{wrapfigure}

Previous studies have shown that, under the flow matching framework, shifting inference timesteps toward high-noise regions, i.e., allocating more integration steps to the early stage, can improve zero-shot TTS performance.
For example, Sway Sampling, introduced in F5-TTS~\citep{F5-TTS}, adopts a cosine-based timestep transformation and controls the degree of shifting through a coefficient $s'$.
Nevertheless, we empirically find that even its strongest shift setting, $s'=-1.0$, remains insufficient for high-dimensional raw waveform generation.
To enable more flexible timestep allocation, we propose PolyShift sampling, which allows a stronger bias toward high-noise regions.
As shown in Figure~\ref{fig:infer_sampling_density}, PolyShift with $p=2.0$ and $s=1.0$, which applies only the polynomial transformation without additional time shifting, already produces a stronger concentration toward $t\rightarrow 0$ than Sway Sampling with $s'=-1.0$.
Increasing the shift factor to $s=3.0$ further moves the sampling density toward the high-noise region, resulting in a substantially higher probability density in the early interval $t\in(0,0.2)$ and a lower density in the low-noise interval $t\in(0.8,1.0)$.
In principle, by jointly adjusting the polynomial factor $p$ and the shift factor $s$, PolyShift can cover a broad range of timestep allocation patterns, making it adaptable to different inference settings.

\section{Diffusion Modeling with STFT and MDCT Representations}
\label{app:frequency_domain_diffusion}

To ensure a fair comparison with direct waveform modeling, we adapt the STFT and MDCT representations to the same flow matching framework with a comparable parameter count.
Specifically, we use the same DiT backbone as the larger configuration described in Appendix~\ref{app:implement_details}.
The model retains the $x$-prediction objective, with the prediction space changed from raw waveforms to STFT or MDCT coefficients. 
Since the prediction targets inherently capture time-frequency information, we omit the auxiliary multi-scale mel-spectrogram loss. 
All other training and inference configurations follow the setup described in Section~\ref{sec:exp_setup}.
Below, we detail the feature transformations and specific modeling adaptations for both representations.

\textbf{Short-Time Fourier Transform (STFT).}
For STFT-based diffusion modeling, we first resample the audio to 16 kHz and extract STFT features using a Hann window, with the FFT size and window length set to 400 and the hop length set to 160. 
This yields a frame rate of 100 Hz, matching that of the patchified waveform representation used in WavTTS. 
To handle the complex-valued STFT coefficients, we separate their real and imaginary parts and concatenate them along the feature dimension, resulting in a 402-dimensional continuous feature sequence (i.e., $2 \times (400 / 2 + 1)$).
Empirical statistics on the Emilia dataset show that the standard deviation of the unrolled STFT coefficients is naturally close to 1, aligning well with the Gaussian noise scale in flow matching. 
We therefore feed the STFT features directly into the diffusion process without additional scaling.
Consistent with the waveform setting, we optimize the model with the MSE loss between the predicted and ground-truth STFT features. 
During inference, once the predicted real and imaginary components are generated via the ODE solver, we reconstruct the final audio waveform using the inverse STFT (iSTFT).

\textbf{Modified Discrete Cosine Transform (MDCT).}
For MDCT-based diffusion modeling, we first resample the audio to 16 kHz and extract MDCT features using a standard Vorbis window with a length of 320.
Due to the overlapped-transform property of MDCT, the hop length is set to half of the window length (i.e., 160), again yielding a frame rate of 100 Hz. 
Under this configuration, each frame corresponds to a 160-dimensional continuous feature vector. 
Empirical statistics on Emilia show that the standard deviation of the MDCT features is approximately 0.005.
We therefore multiply the MDCT coefficients by a factor of 200 to align their scale with the Gaussian noise used in flow matching. 
The scaled MDCT coefficients are then fed into the flow matching process and optimize the model with the MSE loss between the predicted and ground-truth MDCT features. 
During inference, after the ODE solver generates the predicted MDCT features, we reconstruct the final waveform using inverse MDCT (iMDCT).